\documentclass{jnmp}

\def \ie {i.e.~}
\def\RHS{r.h.s.~}
\def \D {\hbox{d}}
\def \Log {\mathop{\rm Log}\nolimits}

\JNMPnumberwithin{equation}{section}

\theoremstyle{definition}

\begin{document}

%
\renewcommand{\evenhead}{Robert Conte}
\renewcommand{\oddhead}
 {Partial integrability of the anharmonic oscillator}

%
\thispagestyle{empty}

\FirstPageHead{2007}{14}{3}{\pageref{firstpage}--\pageref{lastpage}}{Article}

\copyrightnote{2007}{Robert Conte}

\Name
 {Partial integrability of the anharmonic oscillator}

\label{firstpage}

\Author{Robert CONTE}

\Address{Service de physique de l'\'etat condens\'e (CNRS URA 2464,
         CEA--Saclay
\\ F--91191 Gif-sur-Yvette Cedex, France \\
~~E-mail: Robert.Conte@cea.fr\\[10pt]}

\Date{Received January 29, 2007; Accepted in revised form April 9, 2007}

\begin{abstract}
\noindent
We consider the anharmonic oscillator with
an arbitrary-degree anharmonicity,
a damping term and a forcing term,
all coefficients being time-dependent:
\begin{eqnarray}
& &
u'' + g_1(x) u' + g_2(x) u + g_3(x) u^n + g_4(x) = 0, \qquad n \hbox{ real}.
\nonumber
\end{eqnarray}
Its physical applications range from
the atomic Thomas-Fermi model to
Emden gas dynamics equilibria,
the Duffing oscillator and numerous dynamical systems.
The present work is an overview which includes and generalizes
all previously known results of partial integrability of this oscillator.
We give the most general two conditions on the coefficients under which
a first integral of a particular type exists.
A natural interpretation is given for the two conditions.
We compare these two conditions with those provided by the
Painlev\'e analysis.
\end{abstract}

\section{Introduction}
\label{sectionIntroduction}

The harmonic oscillator is the simplest approximation to a physical oscillator
and, when perturbation terms are taken into account,
the resulting \textit{anharmonic oscillator} is governed by
the nonlinear differential equation
\begin{eqnarray}
& &
E \equiv u'' + g_1 u' + g_2 u + g_3 u^n + g_4 = 0,\
n (n-1) g_3 \not=0,
\label{eq1}
\end{eqnarray}
where $'$ denotes the derivative with respect to the independent
time or space variable $x$,
$g_1(x)$ a damping factor,
$g_2(x)$ a time-dependent frequency coefficient,
$g_3(x)$ the simplest possible anharmonic term,
$g_4(x)$ a forcing term.

As to the anharmonicity exponent $n$,
it can be either real if $u(x)$ is real positive,
which is the case for Lane-Emden \cite{Lane} gas dynamics equilibria,
rational,
like $n=3/2$ in the Thomas and Fermi \cite{Thomas,Fermi} atomic model,
or more usually integer:
$-3$ for the Ermakov \cite{Ermakov} or Pinney \cite{Pinney} equation,
$3$ for the Duffing oscillator \cite{Duffing}.

For generic values of the coefficients,
this equation is equivalent to a third order autonomous dynamical system,
which generically admits no closed form general solution.
The purpose of this article is to review all the nongeneric situations
for which there exist exact analytic results,
such as a first integral or a closed form solution,
either particular or general.
This can only happen when the coefficients satisfy some constraints.

The paper is organized as follows.
In Section \ref{sectionLH},
we give a Lagrangian and a Hamiltonian formulation for any value
of the coefficients $(n,g_i)$.
This generalizes all the previous particular results,
obtained for values of
$(n, g_1, g_2, g_3, g_4)$
equal to:
\begin{description}
\item
$(5;0,\hbox{const,const},0)$        \cite{Chandrasekhar},
\item
$(5;2/x,0,1,0)$                     \cite[Eq.~(3.7)]{Logan},
\item
$(n;g_1,0,g_3,0)$                   \cite{SarletBahar,Sarlet,FeixLewis},
\item
$(n;0,\hbox{const},a x^{\alpha},0)$ \cite{Besnardetal},
\item
$(n;g_1,0,1,0)$                     \cite{Leach1985},
\item
$(n;g_1,g_2,g_3,0)$                 \cite{ESC},
                           \cite[Section 6.74, vol.~1]{Kamke}.
\end{description}

In Section \ref{sectionParticularFirstIntegral},
we provide two conditions on $(n,g_i)$ which are sufficient to
ensure the existence of a first integral.

In Section \ref{sectionInterpretation},
we give a natural interpretation of these two conditions.

Finally, in section \ref{sectionPA},
we perform the Painlev\'e analysis of (\ref{eq1}).
Most of this work has already been done
by Painlev\'e and Gambier \cite{GambierThese}.
Indeed,
the ordinary differential equation (ODE) (\ref{eq1}) belongs,
at least for specific values of $n$
and
maybe after a change $u \mapsto u^N$ of the dependent variable $u$ in case $n$
is not an integer,
to the class of second order ODEs which they studied and classified.
However, as opposed to these classical authors,
we do not request the full Painlev\'e integrability of the ODE,
only some partial integrability,
and this requires some additional work.
In particular,
we compute the condition for the absence of
any infinite movable branching,
\ie\ a multivaluedness which occurs at a location depending on the initial
conditions.
Such a condition, like for linear ODEs,
arises from any integer value of the difference of the two Fuchs indices,
whether positive or negative,
and we check that this condition is a differential consequence
of the two conditions for the existence of a particular first integral.
This detailed Painlev\'e analysis of equation (\ref{eq1})
happens to be an excellent example for several features
of Painlev\'e analysis which are most of the time overlooked.

For convenience, we use the notation
\begin{eqnarray}
& &
\Log G_1(x) = \int^x g_1(t) \D t, \qquad
\gamma_3=\Log g_3, \qquad \gamma_4=\Log g_4,
\label{eq2}
\end{eqnarray}
and the convention that function $G_1$ implicitly contains an
arbitrary multiplicative constant;
letter $K$, with or without subscript, denotes an arbitrary constant.
Function $G_1$ frequently occurs, for the way to suppress term $g_1 u'$
in (\ref{eq1}) is to perform the change of function
$ u \to G_1^{-1/2}u$.

\section{Lagrangian and Hamiltonian formulations}
\label{sectionLH}
For every value of $(n,g_i)$, including the logarithmic case $n=-1$,
the anharmonic oscillator can be put in Lagrangian form
\begin{eqnarray}
& &
 \left(\frac{\partial L}{\partial u'} \right)'
     - \frac{\partial L}{\partial u} = 0,
\label{eq3}
\end{eqnarray}
or in Hamiltonian form
\begin{eqnarray}
& &
q'=   \frac{\partial H}{\partial p},\
p'= - \frac{\partial H}{\partial q},
\label{eq4}
\end{eqnarray}
as shown by the explicit expressions for $L,H,q,p$
\begin{eqnarray}
& &
L(u,u',x) =
G_1 \left[u'^2 - 2 g_3 \int_0^u u^n \D u - g_2 u^2 - 2 g_4 u \right]
+ \frac{1}{2} \left( h u^2 \right)',
\label{eq5}
\\
& &
H(q,p,x) =
G_1 \left[u'^2 + 2 g_3 \int_0^u u^n \D u + g_2 u^2 + 2 g_4 u \right]
- \frac{1}{2} h' u^2,
\label{eq6}
\\
& &
q = u,\
p=2 G_1 u' + h u,
\label{eq7}
\end{eqnarray}
in which $h$ is an arbitrary gauge function of $x$.

\section{Particular first integral}
\label{sectionParticularFirstIntegral}

According to Noether theorem, one can find first integrals by looking at the
infinitesimal symmetries of the Lagrangian.
For a detailed review of this Lie symmetries approach to the anharmonic
oscillator, the interested reader can refer to \cite{Euler1997}.
Since the dependence of ODE (\ref{eq1}) in $u$ is rather simple,
let us determine under which conditions on parameters $(n,g_i)$
there exists a particular first integral containing the same kind of terms than
the Hamiltonian
\begin{eqnarray}
& &
 I = f_1 u'^2 + f_2 \int_0^u u^n \D u + f_3 u u' + f_4 u^2 + f_5 u + f_6,
\label{eq8}
\end{eqnarray}
in which the six functions $f_i$ of $x$ are to be determined.

Eliminating $u''$ between $I'$ and $E$, we obtain
\begin{eqnarray}
I'-(2 f_1 u' + f_3 u) E
&
{\hskip -0.4truecm}
\equiv
{\hskip -0.4truecm}
&
   f_2' \int_0^u u^n \D u - g_3 f_3 u^{n+1}
 +(f_2 - 2 g_3 f_1) u^n u' + f_6'
\label{eq9}
\\
&
{\hskip -0.4truecm}
{\hskip -0.4truecm}
&
 +(f_1' + f_3 - 2 g_1 f_1) u'^2
 +(f'_3 + 2 f_4 - g_1 f_3 - 2 g_2 f_1) u u'
\nonumber
\\
&
{\hskip -0.4truecm}
{\hskip -0.4truecm}
&
 +(f'_4 - g_2 f_3 ) u^2
 +(f_5 - 2 g_4 f_1) u'
 +(f'_5 - g_4 f_3) u.
\nonumber
\end{eqnarray}

Out of the nine monomials
$\int_0^u u^n \D u, u^{n+1}, u^n u', u'^2, u^2, u u', u', u,1$,
only eight are linearly independent since $n(n-1)\not=0$,
thus generating eight linear homogeneous differential equations
in six unknowns,
hence generically two conditions on $(n,g_i)$.
Note that, even in the logarithmic case $n=-1$, the first generated equation
is $f_2' - (n+1) g_3 f_3 = 0$.
Functions $f_2$ to $f_6$ are given by
\begin{eqnarray}
& &
f_2= 2 g_3 f_1,
\label{eq10a}
\\
& &
f_3= 2 g_1 f_1 -f_1',
\\
& &
f_4= (g_1^2 + g_2 -g_1')f_1-\frac{3}{2} g_1 f_1' + \frac{1}{2} f_1'',
\\
& &
f_5= 2 g_4 f_1,
\\
& &
f_6 = \delta_{n,-1} \int^x g_3 f_3 \D x,
\label{eq10d}
\end{eqnarray}
and function $f_1$ must be a nonzero solution common to the three linear
equations
\begin{eqnarray}
& &
{\hskip -1.0truecm}
\left[ -2(n+1) g_1 g_3 + 2 g_3' \right] f_1 + (n+3) g_3 f_1' = 0,
\label{eq11a}
\\
& &
{\hskip -1.0truecm}
(-2 g_1 g_2 + 2 g_1 g_1' -g''_1 + g_2') f_1
  +(g_1^2 + 2 g_2 - \frac{5}{2} g_1') f_1'
  - \frac{3}{2} g_1 f_1''
  + \frac{1}{2} f_1''' = 0,
\label{eq11b}
\\
& &
{\hskip -1.0truecm}
(-2 g_1 g_4 + 2 g'_4) f_1 + 3 g_4 f_1' = 0.
\label{eq11c}
\end{eqnarray}

Each above equation can be integrated once,
\begin{eqnarray}
& &
K_1=f_1^{n+3} G_1^{-2n-2} g_3^2,
\label{eq12a}
\\
& &
K_2=f_1^2 G_1^{-2} g_2
+ \int
\Big[\big(
(g_1^2-g_1')f_1-\frac{3}{2}g_1 f_1'+\frac{1}{2}f_1''\big)' f_1 G_1^{-2}\Big]
 \D x,
\label{eq12b}
\\
& &
K_3=f_1^3 G_1^{-2} g_4^2.
\label{eq12c}
\end{eqnarray}

Whatever be $(n,g_i)$,
the function $f_1$ can always be computed from (\ref{eq11b});
depending on $(n,g_4)$, it is also given by
\begin{eqnarray}
n \not= -3:
& & f_1 = G_1^{2(n+1)/(n+3)}
          g_3^{-2/(n+3)}
\label{eq13a}
\\
g_4 \not= 0:
& & f_1 = G_1^{2/3} g_4^{-2/3}
\label{eq13b}
\\
n=-3:
& & f_1 = - g_3^{-1} y^2,\ y=\hbox{ general solution of }
\nonumber
\\
& & \left[g_3^{-1} y^3 (y''-\frac{1}{2} \frac{g_3'}{g_3}y' - g_2 y) \right]'=0
\label{eq13c}
\end{eqnarray}
where the constants $K_1$ and $K_3$ have been absorbed in the
definition of $G_1$.
The only case in which equation (\ref{eq13c}) needs to be considered is
$n=-3,g_4=0$, and its solution can be found in
\cite{Ermakov,GambierThese}
\cite[\P 14.33]{Ince}
\cite{Pinney}
\cite[Eq.~E12]{BureauMI}
\cite{Conte1992a}.

Once $f_1$ is determined,
$f_2$ and $f_5$ are given by (\ref{eq10a}), (\ref{eq10d}),
and $f_3, f_4$ by the three following expressions,
corresponding to cases (\ref{eq13a}), (\ref{eq13b}), (\ref{eq13c})
respectively,
\begin{eqnarray}
& & \frac{f_3}{f_1} =
\left\lbrace
\matrix{
\displaystyle{
\frac{2}{n+3} \left(2 g_1 +\gamma_3' \right)
}
\hfill \cr
\displaystyle{
\frac{2}{3} (2 g_1 + \gamma_4')
}
\hfill \cr
\displaystyle{
- \gamma_3' - \frac{f_1'}{f_1}
}
\hfill \cr}
\right.
\label{eq14a}
\\
& & \frac{f_4}{f_1} =
\left\lbrace
\matrix{
\displaystyle{
g_2 + \frac{
 -2(n-1)g_1^2-(n+3)(2 g_1' + \gamma_3'')-(n-5)g_1 \gamma_3'+ 2 \gamma_3'^2
}{(n+3)^2}
}
\hfill \cr
\displaystyle{
 g_2 +\frac{2}{9} g_1^2-\frac{2}{3} g_1'-g_1 \gamma_4'
 - 2 \gamma_4'^2 - \gamma_4''
}
\hfill \cr
\displaystyle{
g_2
 + \frac{1}{4} \gamma_3'^2
 + \frac{1}{2} \gamma_3''
 + \frac{3}{4} \gamma_3' \frac{f_1'}{f_1}
 + \frac{1}{2} \frac{f_1''}{f_1}.
}
\hfill \cr}
\right.
\label{eq14b}
\end{eqnarray}

Parameters $(n,g_i)$ must satisfy the conditions,
polynomial in $n, g_1, g_2, \gamma_3', \gamma_4'$, resulting from the
elimination of $f_1$ between the three linear equations
(\ref{eq11a})-(\ref{eq11c}).
There are two such conditions when $ g_3 g_4 $ is nonzero,
and only one when it is zero.
The simplest choice of these two conditions is
(the labelling refers to the contributing $g_i$'s):
\begin{eqnarray}
g_4 \not= 0: C_{134} \equiv
& &
2 n g_1 - 3 \gamma_3' + (n+3) \gamma_4' = 0,
\label{eq15a}
\\
g_4 \not= 0: C_{124} \equiv
& &
4 g_1^3
-18 g_1 g_2
-18 g_1''
+ 27 g_2'
+ (6 g_1^2 -36 g_2 + 27 g_1') \gamma_4'
\nonumber
\\
& &
-6 g_1 \gamma_4'^2
-4 \gamma_4'^3
+ 9 g_1 \gamma_4''
+18 \gamma_4' \gamma_4''
-9 \gamma_4''' = 0
\label{eq15b}
\end{eqnarray}
uniquely defined as, respectively, the condition independent of
$g_2$ and the one independent of $(n,g_3)$.
By elimination, one obtains the condition independent of $g_4$ and the one
independent of $g_1$,
\begin{eqnarray}
C_{123} \equiv
& &
{\hskip -0.4 truecm}
-4 \left( 2 g_1 + \gamma_3' \right)^3
+(n+3) [
 (n-3)(-4 g_1^3 -2 g''_1 -4 g_2 \gamma_3' - \gamma_3''')
\nonumber
\\
& &
{\hskip -0.4 truecm}
 +2(n+3)(n-1) g_1 g_2
 +n (-8 g_1 g_1' -6 g_1 \gamma_3'^2)
 +(n+3)^2 g_2'
\nonumber
\\
& &
{\hskip -0.4 truecm}
 +(n-9)(-2 g_1^2 \gamma_3' - g_1'  \gamma_3')
 -3 (n-1) g_1 \gamma_3''
 +6 \gamma_3' \gamma_3''
 ] = 0,
\label{eq15c}
\\
g_4 \not= 0: C_{234} \equiv
& &
{\hskip -0.4 truecm}
n^3 g_2'
+n^2 (-g_2 \gamma_3'- \gamma_3'''+\frac{3}{2} \gamma_3'' \gamma_4'
      +\frac{1}{2} \gamma_3' \gamma_4'' -2 \gamma_4' \gamma_4'' + \gamma_4''')
\nonumber
\\
& &
{\hskip -0.4 truecm}
-\frac{3}{2} \gamma_3'^2 \gamma_4'
+\frac{1}{2} \gamma_3'^3
-n^2(n-1) g_2 \gamma_4'
\nonumber
\\
& &
{\hskip -0.4 truecm}
-\frac{1}{2} (n^2-3) \gamma_3' \gamma_4'^2
+\frac{1}{2} (n^2-1) \gamma_4'^3 = 0.
\label{eq15d}
\end{eqnarray}

For $(n+3) g_4 \not= 0$, any two of the above four conditions are
functionally independent.
For $n=-3$, one has $ 27 C_{123} - 4 C_{134}^3 = 0$
and independent conditions are $C_{134}$ and $C_{234}$.
All above conditions admit an integrating factor,
a natural consequence of the integrated forms
(\ref{eq12a})--(\ref{eq12c}).
This is evident for $C_{134}$;
for each of the three others, it is sufficient to integrate it
as a first order linear inhomogeneous ODE in $g_2$,
\begin{eqnarray}
g_4 \not= 0: K_{134} \equiv
& &
{\hskip -0.5truecm}
G_1^{2n} g_3^{-3} g_4^{n+3},
\label{eq16a}
\\
g_4 \not= 0: K_{124} \equiv
& &
{\hskip -0.5truecm}
 \left[
 g_2
 -\frac{2}{9} g_1^2
 -\frac{2}{3} g_1'
 +\frac{1}{9} g_1 \gamma_4'
 +\frac{1}{9}     \gamma_4'^2
 -\frac{1}{3}     \gamma_4''
 \right]
 G_1^{-8/3} g_4^{-4/3} ,
\label{eq16b}
\\
K_{123} \equiv
& &
{\hskip -0.5truecm}
[(n+3)^2 g_2
 - (n+3) (2 g_1' + \gamma''_3)
 -2(n+1)g_1^2
\nonumber
\\
& &
{\hskip -0.5truecm}
 -(n-1) g_1 \gamma'_3 +\gamma'^2_3]^{n+3}
 G_1^{2(n-1)} g_3^{-4},
\label{eq16c}
\\
g_4 \not= 0: K_{234} \equiv
& &
{\hskip -0.5truecm}
 \left[
 g_2 + \frac{
 (n+2) \gamma_3' \gamma_4'
 -\gamma_3'^2
 -(n+1) \gamma_4'^2
 + 2 n (\gamma_3'' - \gamma_4'')}{2 n^2}
 \right] \times
\nonumber
\\
& &
{\hskip -0.5truecm}
 g_3^{-4/3} g_4^{4/n}.
\label{eq16d}
\end{eqnarray}

In the Duffing case $n=3$, condition $C_{123}$ has already been given
\cite{ESC},
together with its integrated form $K_{123}$ \cite{Estevez1991}.

\section{Interpretation of the two conditions}
\label{sectionInterpretation}

A very simple interpretation can be given for the two conditions.
Indeed,
the form of equation (\ref{eq1}) is invariant under the simultaneous change of
dependent and independent variables
\begin{eqnarray}
& &
 u(x) \to U(X): \qquad u= \alpha(x) U , X=\xi(x),
\label{eq17}
\end{eqnarray}
where $\alpha$ and $\xi$ are two arbitrary gauge functions.
The transformed ODE reads
\begin{eqnarray}
& &
 U''
+\frac{1}{\xi'}
 \left[g_1+2\frac{\alpha'}{\alpha}+\frac{\xi''}{\xi'} \right] U'
+\frac{1}{\xi'^2}
 \left[g_2+g_1\frac{\alpha'}{\alpha}+\frac{\alpha''}{\alpha} \right] U
\nonumber
\\
& &
+\frac{\alpha^{n-1}}{\xi'^2} g_3 U^n
+\frac{1}{\alpha \xi'^2}g_4 = 0.
\label{eq18}
\end{eqnarray}

Let us adjust the two functions $\alpha, \xi$ so as to make two of the
four new coefficients as simple as possible.
One of the three possible ways is to cancel the damping term by the choice
$ \xi' = \alpha^{-2} G_1^{-1}$, which reduces ODE (\ref{eq18}) to
\begin{eqnarray}
& &
 U''
+ \alpha^4 G_1^2
 \left[g_2+g_1\frac{\alpha'}{\alpha}+\frac{\alpha''}{\alpha} \right] U
+\alpha^{n+3} G_1^2 g_3 U^n
+\alpha^3 G_1^2 g_4 = 0.
\label{eq19}
\end{eqnarray}

Canceling the new $g_2$ coefficient amounts to solving the general linear
second order ODE for $\alpha$,
which is possible (from the point of view of Painlev\'e, adopted here)
but does not lead to an explicit value of $\alpha$.
This reduced form is then
\begin{eqnarray}
& &
U'' + h_3 U^n + h_4=0,
\end{eqnarray}
and this means that one can freely set $g_1=g_2=0$ in (\ref{eq1})
without altering its global properties (existence of first integrals,
Painlev\'e property, etc).
Instead of that,
one can make
constant either the reduced $g_3$ coefficient iff $(n+3)g_3 \not= 0$
by choosing $\alpha^{n+3}=G_1^{-2} g_3^{-1}$,
or the reduced $g_4$ coefficient iff $g_4 \not= 0$ by the choice
$\alpha = G_1^{-2/3} g_4^{-1/3}$
(let us recall that $G_1$
implicitly contains an arbitrary multiplicative constant).

We are thus led to the reduced forms
\begin{eqnarray}
g_4 \not= 0:
& &
{\hskip -0.4truecm}
g_1 \mapsto 0,\
g_4 \mapsto 1,\
\label{eq20a}
\\
& &
{\hskip -0.4truecm}
g_2 \mapsto \left[
 g_2
 -\frac{2}{9} g_1^2
 -\frac{2}{3} g_1'
 +\frac{1}{9} g_1 \gamma_4'
 +\frac{1}{9}     \gamma_4'^2
 -\frac{1}{3}     \gamma_4''
 \right]
\times
\nonumber
\\
& &
{\hskip -0.4truecm}
\phantom{g_2 \mapsto }
 G_1^{-8/3} g_4^{-4/3},\
\nonumber
\\
& &
{\hskip -0.4truecm}
g_3 \mapsto g_3 G_1^{-2n/3} g_4^{-(n+3)/3},
\nonumber
\\
n\not=-3:
& &
{\hskip -0.4truecm}
g_1 \mapsto 0,\
g_3 \mapsto 1,\
\label{eq20b}
\\
& &
{\hskip -0.4truecm}
g_2 \mapsto \big[
 g_2
 -\frac{1}{n+3} (2 g_1' + \gamma_3'')
 +\frac{1}{(n+3)^2}
  (-2(n+1)g_1^2
\nonumber
\\
& &
{\hskip -0.4truecm}
\phantom{g_2 \mapsto }
 -(n-1) g_1 \gamma_3' +\gamma_3'^2 )\big]
 G_1^{2(n-1)/(n+3)} g_3^{-4/(n+3)},\
\nonumber
\\
& &
{\hskip -0.4truecm}
g_4 \mapsto g_4 G_1^{2n/(n+3)} g_3^{-3/(n+3)},
\nonumber
\\
n=-3, g_4=0:
& &
{\hskip -0.4truecm}
g_1 \mapsto 0,\
g_3 \mapsto g_3 G_1^2,\
\nonumber
\\
& &
{\hskip -0.4truecm}
g_2 \mapsto
 \left[g_2+g_1\frac{\alpha'}{\alpha}+\frac{\alpha''}{\alpha} \right]
 \alpha^4 G_1^{2} \mapsto 0.
\label{eq20c}
\end{eqnarray}

Then the interpretation is obvious:
any reduced coefficient distinct from $0$ or $1$
is the \RHS\ of one of the integrated conditions (\ref{eq16a})--(\ref{eq16d}).
Conversely, any integrated condition is one of the remaining
coefficients when two coefficients have been made constant by a choice of
gauge.
For instance, $K_{234}$ is the reduced $g_2$ coefficient associated to
reduced coefficients $g_3$ and $g_4$ equal to unity.

This can also be seen in a more elementary way.
In a gauge $(\alpha, \xi)$ such that $g_1=0, g_3' g'_4=0$,
an expression for the first integral is
\begin{eqnarray}
& &
 g_1=0, g_3' g'_4=0:
 I_0= u'^2 + 2 g_3 \int_0^u u^n \D u + g_2 u^2 + 2 g_4 u,
\label{eq21}
\end{eqnarray}
and, from the relation
\begin{eqnarray}
& &
 I'_0-2 u' E \equiv
 2 g_3' \int_0^u u^n \D u + g_2' u^2 + 2 g'_4 u,
\label{eq22}
\end{eqnarray}
one deduces that the two other coefficients $g_2$ and $g_3$ or $g_4$
must be constant.

The Hamiltonian (\ref{eq6}) is a first integral if and only if
$g_1=0$ and all other $g_i$'s are constant.

\section{Painlev\'e analysis}
\label{sectionPA}

Painlev\'e\ set up the problem of finding nonlinear differential equations
able to define functions,
just like the first order elliptic equation
\begin{eqnarray}
& &
u'^2=4 u^3 - g_2 u - g_3,\
(g_2,g_3) \hbox{ complex constants},
\end{eqnarray}
defines the elliptic function of Weierstrass $\wp(x,g_2,g_3)$,
a doubly periodic function which includes as particular cases the
well known trigonometric and hyperbolic functions.
For a tutorial introduction,
see the books \cite{Hille,Cargese96}.

A by-product of this quest for new functions has been the construction
of exhaustive lists of nonlinear differential equations,
the general solution of which can be made singlevalued
(in more technical terms, without movable critical singularities,
this is the so-called \textit{Painlev\'e property} (PP)),
which implies that their general solution is known in closed form.
In particular,
the list of second order first degree algebraic equations, i.e.
\begin{eqnarray}
& &
u''=F(u',u,x),
\label{eqGambierclass}
\end{eqnarray}
with $F$ rational in $u'$, algebraic in $u$, analytic in $x$,
which possess the PP
has been established by Painlev\'e and Gambier \cite{GambierThese}.

These classical results apply to our problem
only for those values of $n$ for which Eq.~(\ref{eq1}),
maybe after a monomial change of the dependent variable $u=U^k,k \in {\cal R}$,
belongs to the class (\ref{eqGambierclass}).
These values, which include at least all the integers, are determined below.
Then, the way those classical results can be applied is twofold.
\begin{enumerate}
\item
Require the PP for our equation or its transform under $u=U^k$.

\item
Restricting to the values of $(n,g_i)$ for which the first integral
(\ref{eq8}) exists,
check that the two conditions for the existence of this first integral
imply the identical satisfaction of
the necessary condition that Eq.~(\ref{eqGambierclass}) have no
movable logarithmic branch points.
Indeed,
this is a classical result of Poincar\'e that the movable singularities
(\ie\ those which depend on the initial conditions)
of first order algebraic ODEs
can only be algebraic,
\ie\ $u \sim u_0 (x-x_0)^p$,
and never logarithmic, \ie\ with some $\Log (x-x_0)$ term.
Let us do that without too many technical considerations.

\end{enumerate}

The above mentioned necessary condition that Eq.~(\ref{eqGambierclass}) have
no movable logarithmic branch points can only be computed
after performing the following steps
(for the unabribged procedure, see \cite[section 6.6]{Cargese96Conte}).

\textit{Step 1}.
For each \textit{family of movable singularities}
\begin{eqnarray}
& &
u=\chi^p \sum_{j=0}^{+\infty} u_j \chi^j,\
u_0\not=0,\
\chi'=1,
\label{eqLaurentu}
\end{eqnarray}
determine the \textit{leading behaviour} $(p,u_0)$.
This is achieved by balancing the highest derivative $u''$
with a nonlinear term.
Therefore, there exist two leading behaviours,
denoted
``family $g_3$'' (balancing of $u''$ and $g_3 u^n$)
and
``family $g_4$'' (balancing of $u''$ and $g_4$)
\begin{eqnarray}
(g_3) \ : \
& &
p=-\frac{2}{n-1},\
u_0=\left[-2 \frac{n+1}{(n-1)^2} g_3 \right]^{\displaystyle{\frac{1}{n-1}}},\
n\not=-1,\
\\
(g_4) \ : \
& &
p=2,\
u_0=-\frac{1}{2} g_4,\
g_4 \not=0.
\end{eqnarray}

\textit{Step 2}.
For each family,
compute the Fuchs indices,
\ie\ the roots $i$ of the indicial equation of the linear equation
obtained by linearizing (\ref{eq1}) near its leading behaviour
$u \sim u_0 \chi^p$,
and require every Fuchs index to be integer.
This linearized equation is
\begin{eqnarray}
(g_3) \ : \
& &
\left(\frac{\D^2}{\D x^2} + n g_3 (u_0 \chi^2)^{n-1}\right) v=0,
\\
(g_4) \ : \
& &
\left(\frac{\D^2}{\D x^2} + 0\right) v=0,
\end{eqnarray}
and the Fuchs indices are obtained by requiring the solution
$v=v_0 \chi^{p+i} (1 + O(\chi))$,
\begin{eqnarray}
(g_3) \ : \
& &
(i+1)\left(i-2-\frac{4}{n-1}\right)=0,
\\
(g_4) \ : \
& &
(i+1)(i+2)=0.
\end{eqnarray}
The diophantine condition that $i=2+4/(n-1)$ be integer has a countable
number of solutions since we have not yet put restrictions on $n$.

\textit{Step 3}.
For each family,
compute all the necessary conditions for the absence of movable logarithms
(in short, no-log conditions),
which might occur when one computes the successive coefficients $u_j$ of
(\ref{eqLaurentu}).
One can check that the family $g_4$ can never generate such no-log
conditions.
These conditions need not be computed on the original equation (\ref{eq1}),
they can be computed on any algebraic transform if this proves more convenient
(indeed, movable logarithms are not affected by an algebraic transform on $u$),
such as
\begin{eqnarray}
& &
u=U^k\ : \
k U U'' + k(k-1) U'^2 + k g_1 U U' + g_2 U^2
\nonumber
\\
& &
\phantom{xxxxxxxxxx}
+ g_3 U^{2+(n-1)k} + g_4 U^{2-k}=0.
\label{eq125}
\end{eqnarray}
The transformed powers $p$ are $p_3=-2/((n-1)k), p_4=2/k$,
and the Fuchs indices are unchanged.

The computation of the no-log conditions is impossible
unless there exists a $k$ making all the powers of $U$ in (\ref{eq125})
at least rational.
In order to avoid the technical complications of dealing with rational
values of the leading exponent $p$,
we restrict to those values of $n$ for which there exists a $k$
making $2+(n-1)k$ and, if $g_4$ is nonzero, $2-k$ integer.
The useful transforms are
\begin{eqnarray}
u=v^{2/(n-1)}:
& &
 \frac{2}{n-1} v v'' -2 \frac{n-3}{(n-1)^2} v'^2 + \frac{2}{n-1} g_1 v v'
 + g_2 v^2
\nonumber
\\
& &
 + g_3 v^4 + g_4 v^{2-2/(n-1)}=0,
\label{eq126a}
\label{eq25}
\\
u=w^{1/(n-1)}:
& &
 \frac{1}{n-1} w w'' - \frac{n-2}{(n-1)^2} w'^2 + \frac{1}{n-1} g_1 w w'
 + g_2 w^2
\nonumber
\\
& &
+ g_3 w^3 + g_4 w^{2-1/(n-1)}=0,
\label{eq126b}
\\
u=V^{-2}:
& &
 - 2 V V'' + 6 V'^2 - 2 g_1 V V'
 + g_2 V^2
\nonumber
\\
& &
+ g_3 V^{4 - 2 n} + g_4 V^4=0,
\label{eq126c}
\\
u=W^{-1}:
& &
 - W W'' + 2 W'^2 - g_1 W W'
 + g_2 W^2
\nonumber
\\
& &
+ g_3 W^{3 - n} + g_4 W^3=0,
\label{eq126d}
\end{eqnarray}
which are polynomial if and only if
\begin{eqnarray}
(\ref{eq126a}):
& &
g_4=0 \hbox{ or } (g_4 \not=0 \hbox{ and } \frac{2}{n-1} \in {\mathcal Z}),
\label{eq127a}
\\
(\ref{eq126b}):
& &
g_4=0 \hbox{ or } (g_4 \not=0 \hbox{ and } \frac{1}{n-1} \in {\mathcal Z}),
\label{eq127b}
\\
(\ref{eq126c}):
& &
2n \in {\mathcal Z},
\label{eq127c}
\\
(\ref{eq126d}):
& &
 n \in {\mathcal Z}.
\label{eq127d}
\end{eqnarray}
The original ODE (\ref{eq1}) is identical to (\ref{eq126a}) for $n=3$
and to (\ref{eq126b}) for $n=2$.

To summarize, let us compute the no-log condition $Q_i=0$ on the ODE for $v$
(\ref{eq126a}).
Unfortunately,
one does not know how to obtain the dependence of $Q_i$ on $n$,
since $n$ must first be given a numerical value before $Q_i$ is computed;
this makes uneasy the comparison with conditions
(\ref{eq15a})--(\ref{eq15b}), which depend on $n$.

To fix the ideas, a list of useful values of $(n,i)$ is displayed in Table
\ref{Table1}.
\tabcolsep=1.5truemm
\tabcolsep=0.5truemm
\tabcolsep=0.8truemm

\begin{table}[h]
\caption[garbage]{Values of $(i,n)$ for $i$ integer $\in [-4,10]$.}
\vspace{0.1truecm}
\begin{center}
\begin{tabular}{|c| c | c | c | c | c | c | c | c | c | c | c | c | c | c |c|}
\hline
$2 + \displaystyle{\frac{4}{n-1}}$
& -4 & -3 & -2 & -1 & 0 & 1 & 2 & 3 & 4 & 5 & 6 & 7 & 8 & 9 & 10
\\
\hline
$n$
& 1/3 & 1/5 & 0 & -1/3 & -1 & -3 & $\infty$ & 5 & 3 & 7/3 & 2 & 9/5 & 5/3
& 11/7 & 3/2
\\
\hline
\end{tabular}
\end{center}
\label{Table1}
\end{table}

The computation of $Q_i$ for positive values of $i$ is classical
\cite{GambierThese,BureauMI}.
Denoting for shortness $C_1=C_{123}, C_2=C_{134}$,
one finds
the following expressions $Q_i$ for the indicated values of
$(n,g_4)$,
\begin{eqnarray}
(-3,0): Q_1=
& &
{\hskip -0.4 truecm}
C_1,
\label{eq26.1}
\\
( 5,0): Q_3=
& &
{\hskip -0.4 truecm}
C_1,
\label{eq26.3}
\\
( 3,g_4): Q_4=
& &
{\hskip -0.4 truecm}
\pm \frac{1}{864} [-2 g_3]^{-1/2} [(\gamma_3' - g_1) C_1 - C_1']
 + \frac{1}{6} C_2,
\label{eq26.4}
\\
(\frac{7}{3},0): Q_5=
& &
{\hskip -0.4 truecm}
 (. g_1^2 + . g_2 + . g_1' + . g_1 \gamma'_3 + . \gamma_3'^2) C_1
 + (. g_1 + . \gamma_3') C'_1 + . C''_1,
\label{eq26.5}
\\
(2, g_4): Q_6=
& &
{\hskip -0.4 truecm}
 . C_1 + . C_1^2 + . C'_1 + . C''_1 + . C'''_1 + . C_2 + . C'_2,
\label{eq26.6}
\\
(\frac{9}{5},0): Q_7=
& &
{\hskip -0.4 truecm}
. C_1 + \dots + . C_1^{(4)},
\label{eq26.7}
\\
(\frac{5}{3},0): Q_8=
& &
{\hskip -0.4 truecm}
. C_1 + \dots + . C_1^{(5)}.
\label{eq26.8}
\end{eqnarray}
where dots stand for rational numbers when $i=5$ and polynomials of
$g_1$, $g_2$, $\gamma_3$, $\gamma_4$ when $i>5$.
Similar relations have been checked for $i=9$ and $i=10$ (Thomas-Fermi case)
but are not reproduced here.
Condition $Q_4=0$ contains a $\pm$ sign arising from the two possible choices
for $v_0$ and is equivalent to the two conditions
\begin{eqnarray}
& &
 (\gamma_3'- g_1) C_{123} - C_{123}' = 0, \qquad C_{134} = 0.
\label{eq27}
\end{eqnarray}

We therefore check the property that each $Q_i$
is indeed a differential consequence of the two
conditions $C_{123}=0, C_{134}=0$ for the existence of a first integral
(\ref{eq8})
\begin{eqnarray}
& &
 \forall i \in {\mathcal N}, \forall g_i: (C_1=0, C_2=0) \Rightarrow (Q_i=0).
\label{eq28}
\end{eqnarray}

For negative \cite{FP1991,CFP1993} values of the Fuchs index $i$,
the results \cite{CFP1993}
are the following:
the family $g_4$ never generates any no-log condition,
and, for the family $g_3$,
a no-log condition arises from the Fuchs index $-1$,
and this condition is a differential consequence of conditions
(\ref{eq15a})--(\ref{eq15b}),
at least for the examples handled
$(n,r,g_4)=(1/5,-3,0), (1/3,-4,0)$.
This is also an experimental verification of
\begin{eqnarray}
& &
 \forall i \in {\mathcal Z}, \forall g_i: (C_1=0, C_2=0) \Rightarrow (Q_{-1}=0)
\label{eq29}
\end{eqnarray}
and this relation cannot be reversed, as proven by Painlev\'e and Gambier.
For instance, in the case of the Duffing oscillator
$(n,i,g_4)=(3,4,g_4)$, condition $Q_4=0$ implies the reducibility of $v$ to
the second Painlev\'e transcendent
whereas the stronger conditions $C_1=0, C_2=0$ imply the reducibility of $v$
to an elliptic function.

\textit{Remark}.
When one includes the contribution of the Schwarzian
in the definition of the gradient of the expansion variable $\chi$,
as done in the invariant Painlev\'e analysis \cite{Conte1989},
\begin{eqnarray}
& &
\chi'=
1 + \frac{S}{2} \chi^2,
\label{eq35a}
\end{eqnarray}
all the computed no-log conditions $Q_i=0$,
equations (\ref{eq26.1})--(\ref{eq26.8}),
are independent of this Schwarzian $S$,
as opposed e.g.~to the Lorenz model \cite{CM1991}.
This certainly indicates some hierarchy between the level of
nonintegrability of these two dynamical systems.

\textit{Remark}.
For some small values of $|i|$, there is equivalence
between the no-log condition and (\ref{eq15a})--(\ref{eq15b}).
This nongeneric situation occurs only for
the following values of $(n,i,g_4)$,
\begin{description}
\item
$(-3,1,0)$, i.e. the Ermakov-Pinney equation \cite{Ermakov,Pinney},

\item
$(5,3,0)$, i.e. an equation considered by
Lane and Emden \cite{Lane,Emden},
Chandrasekhar \cite{Chandrasekhar} and Logan \cite[p.~52]{Logan},

\item
$(1/5,-3,0)$, an equation which could deserve more study.
\end{description}

\section{Conclusion}
\indent

This work generalizes all previous results on the partial integrability
of the anharmonic oscillator.
It gives a natural interpretation of the two conditions for the existence
of a particular first integral, in terms of reduced coefficients.
Finally, this system is an excellent example to study
several features of Painlev\'e analysis.

A good, recent bibliography can be found in Ref.~\cite{GoennerHavas}.

\section*{Acknowledgments}
\indent

The author wishes to thank M.~Musette for fruitful discussions
during the completion of this work
and for her encouragement to publish these lecture notes,
a first draft of which was delivered at a meeting in Dijon
\cite{DijonProceedings}.

He thanks the IUAP Contract No.~P4/08 funded by the Belgian government
and
acknowledges the financial support of the Tournesol grant
T99/040.



\label{lastpage}

\vfill \eject

\end{document}